\documentclass[journal=nalefd,manuscript=letter]{achemso}

\usepackage[version=3]{mhchem} 
\usepackage{color}

\author{Anders Pors}
\email{alp@iti.sdu.dk}
\altaffiliation{These authors contributed equally to this work}
\author{Michael G. Nielsen}
\altaffiliation{These authors contributed equally to this work}
\author{Sergey I. Bozhevolnyi}
\affiliation{Department of Technology and Innovation, University of Southern Denmark, Niels Bohrs All{\'{e}} 1, DK-5230 Odense M, Denmark}

\title[Analog computing using plasmonic metasurfaces]
{Analog computing using reflective plasmonic metasurfaces}

\begin{document}
\begin{abstract}
Motivated by the recent renewed interest in compact analog computing using light and metasurfaces (Silva, A. \emph{et al}., Science 2014, 343, 160-163), we suggest a practical approach to its realization that involves reflective metasurfaces consisting of arrayed gold nanobricks atop a subwavelength-thin dielectric spacer and optically-thick gold film, a configuration that supports gap-surface plasmon resonances. Using well established numerical routines, we demonstrate that these metasurfaces enable independent control of the light phase and amplitude, and design differentiator and integrator metasurfaces featuring realistic system parameters. Proof-of-principle experiments are reported along with the successful realization of a high-quality poor-man's integrator metasurface operating at the wavelength of 800\,nm.
\end{abstract}


In the quest to fully control light at the nanoscale, the year 2000 marks a new epoch in studying light-matter interactions, as researchers, fascinated by the experimental verification of negative refraction\cite{smith} and the theoretical work on perfect lensing\cite{pendry2}, in copious amounts ventured into the field of man-made materials, i.e., metamaterials. More than a decade later, several groundbreaking applications have been suggested and verified, such as super-resolution imaging\cite{fang}, invisibility cloaks\cite{pendry}, and metamaterial nanocircuits\cite{engheta}. In any case, however, and especially at optical frequencies, the general usage of metamaterials seem hindered by difficulties in fabrication and too high losses. As a way to circumvent the drawbacks of metamaterials, the two-dimensional analog, known as metasurfaces, have attracted increasing attention in recent years\cite{yu3}. Metasurfaces are characterized by a subwavelength thickness in the direction of propagation, while the transverse plane typically consists of an array of metallic scatterers with subwavelength periodicity. Generally speaking, metasurfaces function as interface discontinuities which, depending on size, shape and composition of scatterers, allow for an abrupt change in the amplitude and/or phase of the impinging light\cite{liu2014}. It should be noted that a single layer of scatterers (due to their Lorentzian-shaped polarizability) only allow for full 2$\pi$-phase control of the cross-polarized light component\cite{yu}, meaning that such metasurfaces have a theoretical efficiency of maximum 25\%\cite{monticone}, though most realizations show efficiencies of a few percent\cite{aieta,aieta2}.

\begin{figure}[H]
	\centering
		\includegraphics[width=8cm]{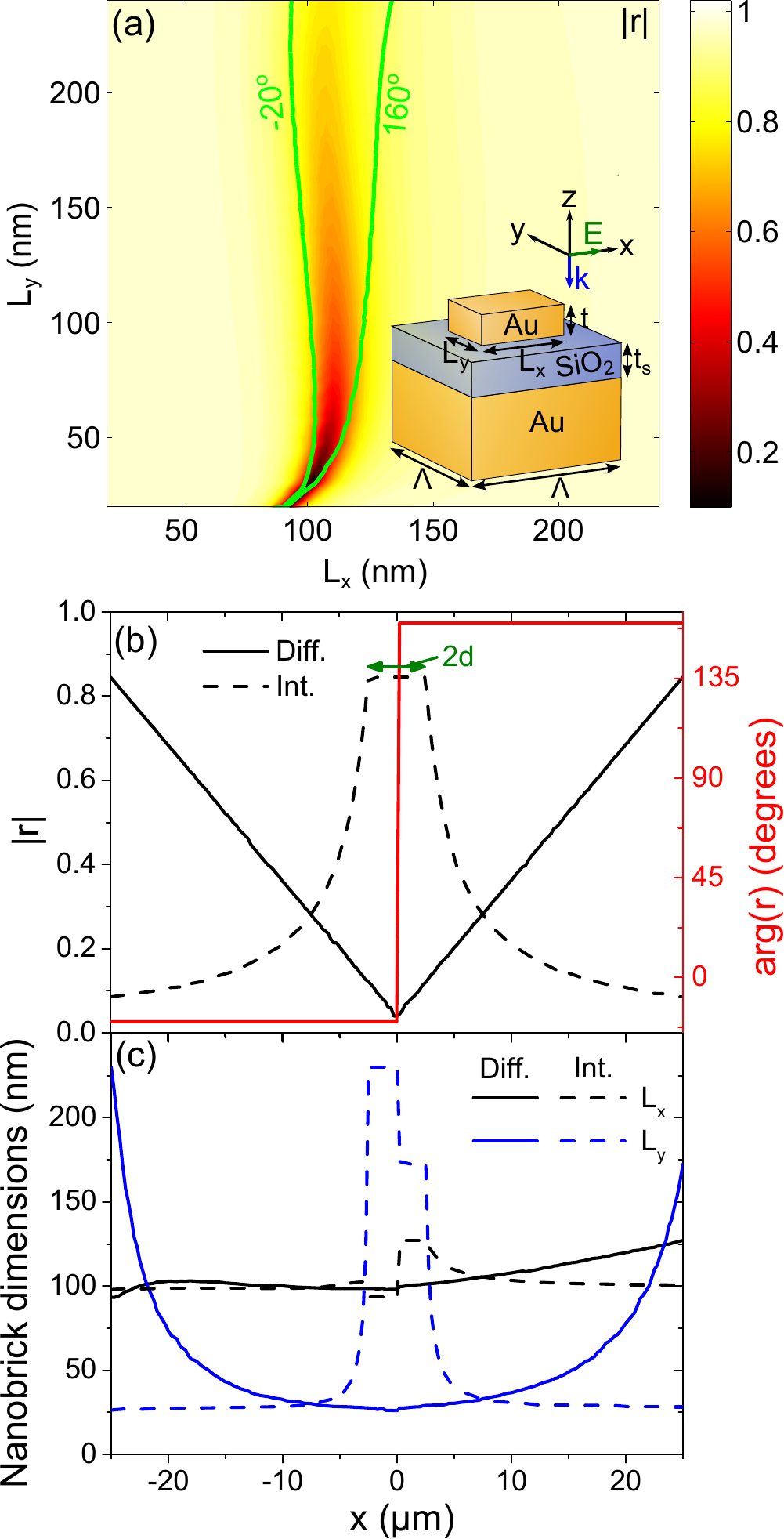}
	\caption{Design of differentiator and integrator metasurfaces. (a) Calculated reflection coefficient $r$ as a function of nanobrick widths for periodic structure consisting of an array of gold nanobricks on top of a glass spacer and gold substrate (see unit cell in inset). The parameters are $\Lambda=250$\,nm, $t=t_s=30$\,nm, $\lambda=800$\,nm, and the incident $x$-polarized light propagates  normal to the surface. Color map shows the reflection coefficient amplitude, while lines are two contours of reflection phase with $\pi$-phase difference. Position-dependent (b) reflection coefficient and (c) nanobrick dimensions of 50\,$\mu$m-wide differentiator and integrator ($d=2.5$\,$\mu$m) metasurfaces. }
	\label{fig:2}
\end{figure}

In order to improve the efficiency of plasmonic metasurfaces, the low-frequency concept of transmit- and reflectarrays has been generalized and adopted to the visible and infrared regimes, where metasurfaces working in transmission consist of several layers in order to reach full phase control and proper impedance matching with surroundings\cite{monticone,pfeiffer}. Accordingly, such metasurfaces are quite complex to fabricate at near-infrared and visible frequencies, with a moderate efficiency of $\sim 20-50$\% due to Ohmic losses in the metal\cite{pfeiffer2,pfeiffer3}. A different approach that works in reflection, which is easy to fabricate and show efficiencies up to $\sim 80$\% for visible light, consists of a periodic arrangement of metal nanobricks on top of a sub-wavelength thin dielectric spacer and optically thick metal film (see inset in Figure \ref{fig:2}a). The full phase and amplitude control of the reflected light are reached by the excitation of gap-surface plasmons (GSPs) that propagate in the gap between the metal film and nanobricks, hence experiencing Fabry-Perot-like resonances due to multiple reflections at nanobrick boundaries\cite{pors2}. As the GSP mode becomes increasingly confined to the gap for decreasing spacer thickness, it is clear that strong modulation in reflection amplitude can be reached, both spectrally and spatially, by a proper variation in nanobrick sizes along the metasurface, allowing one to design broadband super-absorbers\cite{nielsen} or surfaces for color printing with subwavelength resolution\cite{roberts2}. In the other regime of weakly confined GSP modes, the metasurface remains reflective at and around the GSP resonance wavelength despite strong variation in reflection phase, thus permitting the construction of efficient wave plates\cite{hao,pors6,pors7,dai}, focusing mirrors\cite{li,pors5}, blazed gratings\cite{sun2}, and unidirectional surface wave couplers\cite{sun}. Note that the latter two functionalities are obtained by varying the reflection phase linearly along the metasurface (keeping the reflection amplitude close to one and constant), whereas flat focusing mirrors require a parabolic phase profile. More importantly, GSP-based birefringent metasurfaces may be used to independently manipulate orthogonal polarizations, that being either in the context of polarization beam-splitters\cite{farahani,pors3}, surface wave excitation\cite{pors4}, or holography\cite{chen}.

In the above mentioned applications of gradient (i.e., inhomogeneous) metasurfaces the considered functionalities are based on either position-dependent reflection amplitude \emph{or} phase. However, within the important topic of light-based compact analog computing, metasurfaces performing mathematical operations have to exhibit position-dependent amplitude \emph{and} phase response\cite{farahani2,silva}. Note that metasurfaces that would enable independent control of the light phase and amplitude have so far not been realized in the optical domain\cite{liu2014}. The question then naturally arises whether GSP-based metasurfaces, despite their simplicity, could also represent an elegant way to design computing metasurfaces that allow for realization in the optical regime. The answer is 'yes', as will be confirmed in this Letter. Here, we design and verify by numerical calculations GSP-based metasurfaces that perform differentiation and integration at the light wavelength $\lambda=800$\,nm, while proof-of-concept experiments demonstrate the feasibility of realization.  

\begin{figure}[H]
	\centering
		\includegraphics[width=8cm]{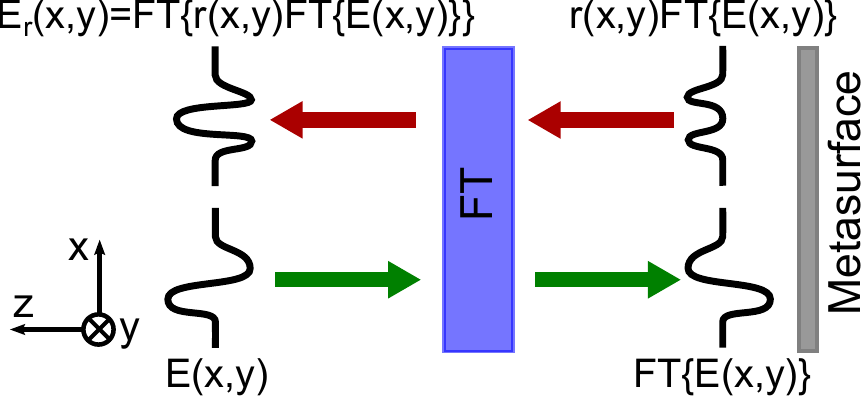}
	\caption{Sketch of system under study which consists of a Fourier transforming (FT) block and a reflective metasurface with position-dependent reflection coefficient $r(x,y)$.}
	\label{fig:1}
\end{figure}

Let us start by reviewing the general idea of computing metasurfaces which is based on the mathematical similarity between convolution in Fourier space (between the system's impulse response and input function) and monochromatic wave interaction with metasurface, together with the Fourier-transforming property of lenses. For a linear space invariant system, described by the two-dimensional impulse response $g(x,y)$, the output $w(x,y)$ for an arbitrarily input function $f(x,y)$ is given by the convolution $w(x,y)=\int g(x-x',y-y')f(x',y')dx'dy'$ or, equivalently,
\begin{equation}
w(x,y)=\mathrm{IFT}\left\{G(k_x,k_y)\mathrm{FT}\left\{f(x,y)\right\}\right\},
\label{eq:conv}
\end{equation}      
where (I)FT means (inverse) spatial Fourier transform, and $G(k_x,k_y)=\mathrm{FT}\{g(x,y)\}$ with ($k_x$,$k_y$) being the spatial frequency variables.
%
%
Limiting our discussion to the system sketched in Figure \ref{fig:1}, it is evident that the expression describing the reflected field $E_r(x,y)$,
\begin{equation}
E_r(x,y)=\mathrm{FT}\left\{r(x,y)\mathrm{FT}\left\{E(x,y)\right\}\right\},
\label{eq:refl}
\end{equation}
is mathematically related to equation \ref{eq:conv}, with the incident field $E(x,y)$ being the input function, the real-space coordinates $(x,y)$ at the metasurface represent ($k_x$,$k_y$), and the position-dependent reflection coefficient $r(x,y)$ is $G(k_x,k_y)$. Note that the system in Figure \ref{fig:1} only includes a single FT-block, which in typical setups represent a regular lens or, for compactness, a graded-index lens\cite{farahani2,silva} or even a focusing metasurface\cite{lin2010}. In any case, the output will, in comparison to equation \ref{eq:conv}, be $w(-x,-y)$ due to the application of FT twice.

In order for the system in Figure \ref{fig:1} to perform mathematical operations the reflection coefficient $r(x,y)$ must mimic the form of the operation in Fourier space. For example, one-dimensional spatial differentiation $\partial/\partial x$ transforms to $ik_x$ in Fourier space, meaning the appropriate $r(x,y)=r(x)$ will be
\begin{equation}
r_\mathrm{diff}(x)=r_\mathrm{m}x/L,
\label{eq:diff}
\end{equation}
where $-L\leq x\leq L$, $2L$ is the size of the metasurface, and $r_\mathrm{m}\leq 1$ is the maximum achievable reflection amplitude. It should be noted that the limited value of the reflection amplitude (between $\sim0$ and $r_\mathrm{m}$) illustrates the fact that calculus-metasurfaces output scaled functions compared to the exact mathematical operation. In the case of one-dimensional integration, described by $(ik_x)^{-1}$ in Fourier space, one needs to handle the singularity at $k_x=0$ when designing the corresponding metasurface. Following the previous suggested approach\cite{silva}, we implement the reflection coefficient
\begin{equation}
r_{\mathrm{int}}(x)=
\begin{cases}
r_\mathrm{m} &, |x|<d \\
r_\mathrm{m}d/ x &, |x|\geq d
\end{cases},
\label{eq:int}
\end{equation}   
where $d \ll L$ defines a region near the center of the metasurface with constant reflectivity $r_\mathrm{m}$.

In an attempt to realize equations \ref{eq:diff} and \ref{eq:int} with easy-to-fabricate metasurfaces, we first numerically study the reflection from GSP-based \emph{homogeneous} metasurfaces at a wavelength of $\lambda=800$\,nm with fixed periodicity $\Lambda=250$\,nm, silicon dioxide (SiO$_2$) spacer thickness $t_s=30$\,nm, and gold nanobrick height $t=30$\,nm (see inset of Figure \ref{fig:2}a). 
%
%
The calculations are performed using the commercial finite-element software Comsol Multiphysics, with the permittivity of gold and SiO$_2$ described by interpolated experimental values\cite{johnson} and the constant $\varepsilon_{\mathrm{SiO}_2}=2.1$, respectively. As seen in Figure \ref{fig:2}a, the carefully chosen geometrical parameters allow for a strong variation in the reflection amplitude near the GSP-resonance for a normal incident $x$-polarized plane wave when varying the widths ($L_x$ and $L_y$) of the nanobrick. Moreover, it is clear that the two contour lines of the reflection phase, with a $\pi$-phase difference, intersect a large span of the reflection amplitude variation due to a weakening of the GSP-resonance for increasing nanobrick size in the direction perpendicular to excitation (i.e., increase in $L_y$). Accordingly, by the assumption that the interaction between neighboring nanobricks is weak, a fact that has been verified in previous studies of GSP-based metasurfaces\cite{pors3}, we can design \emph{inhomogeneous} calculus metasurfaces, defined by equations \ref{eq:diff} and \ref{eq:int}, from the reflection map in Figure \ref{fig:2}a. Here, we set $L=25$\,$\mu$m, $r_\mathrm{m}=0.84$, and $d=2.5$\,$\mu$m, which leads to the position-dependent reflection coefficients and nanobrick dimensions in Figure \ref{fig:2}b and \ref{fig:2}c, respectively. It should be noted that these figures are constructed by following the reflection amplitude along the two phase-contour lines, each representing half of the metasurface, while minimizing $\left||r_\mathrm{ideal}(p\Lambda)|-|r(L_x,L_y)|\right|$ for every nanobrick position $x=p\Lambda$, where $-100\leq p \leq 100$ is an integer, $r_{\mathrm{ideal}}$ corresponds to either equation \ref{eq:diff} or \ref{eq:int}, and $r(L_x,L_y)$ is from contour lines in Figure \ref{fig:2}a. By examination of Figures \ref{fig:2}b and \ref{fig:2}c, it is evident that one can achieve (almost) perfect reflection profiles at the expense of extreme fabrication tolerances on the nanobrick dimensions of $\sim 1$\,nm. The apparent strong sensitivity of reflection to the nanobrick dimensions, especially prominent in the low-reflecting part of the metasurfaces, arises due to the relatively small ($L_x$,$L_y$)-region of high absorption together with concurrent narrowing of phase-contour lines (see Figure \ref{fig:2}a), making the designed metasurfaces, at first glance, difficult to fabricate. Noting, however, that the performance of calculus metasurfaces is only weakly dependent on the exact reflection phase in the low-reflecting part of the metasurfaces, makes it manageable to realize current designs with, e.g., electron beam lithography.

\begin{figure}[H]
	\centering
		\includegraphics[width=15cm]{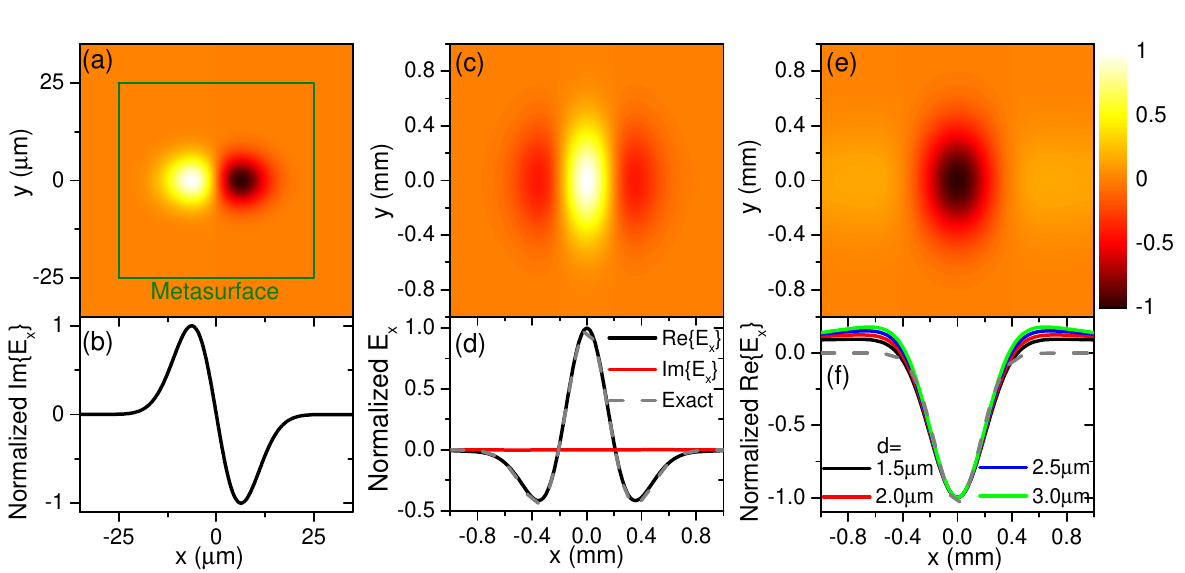}
	\caption{Numerical performance of differentiator and integrator metasurfaces. (a) Incident field on the metasurface representing the Fourier transform of the object function $ax\exp{(-x^2/b-y^2/c)}$ with $a,b,c$ being positive constants. (c,e) Normalized electric far-field (evaluated 10\,mm away from metasurface) from differentiator and integrator metasurfaces, respectively, obtained by electric point-dipole calculations. (b,d,f) represent center-line cross cuts in a,c, and e, respectively, with dashed lines in d,f corresponding to exact solutions.}
	\label{fig:3}
\end{figure}

It is worth noting that despite the seemingly continuous reflection profiles in Figure \ref{fig:2}b, the reflection is only controlled along the $x$-direction in integer steps of $\Lambda$, corresponding to the positions of the nanobricks. Additionally, as already mentioned, the integrator metasurfaces only approximate the operation for small $k_x$-values. In order to better judge on the influence of those imperfections in metasurface performance, we turn to numerical calculations of realistic $50\times50$\,$\mu$m$^2$ metasurfaces consisting of $\sim40,000$ nanobricks. Since such large metasurfaces are too computationally demanding in our current hardware setup, we resort to simpler point-dipole calculations\cite{pfeiffer4} in which each nanobrick is modeled as an electric dipole with dipole moment $\mathbf{p}(x,y)\propto r(x)\mathrm{FT}\{E(x,y)\}\hat{\mathbf{x}}$, with the electric far-field representing the reflected field from the metasurfaces. For ease of comparison with previous work on computing-metasurfaces\cite{silva}, we study a $x$-polarized incident wave with spatial smooth variation $E(x,y)=ax\exp{(-x^2/b-y^2/c)}$, where $a,b,c$ are positive constants. The FT-field incident on the metasurfaces in the following calculations is depicted in Figures \ref{fig:3}a and \ref{fig:3}b, thus demonstrating that parameters are chosen so that practically all spatial frequency components are within the size of the metasurface. 
%
%
Regarding the differentiator metasurface, Figure \ref{fig:3}c displays the normalized electric far-field in the $xy$-plane (evaluated at a distance of $z=10$\,mm away from the metasurface), with the corresponding center-line cross-cut shown in Figure \ref{fig:3}d. Note that for ease of visualization the electric field has been multiplied by a phase factor $\exp(i\phi)$ to ensure a pure real field at the evaluation plane (see Figure \ref{fig:3}d). That said, the designed metasurface shows excellent performance, demonstrating a reflected field whose spatial variation is in perfect agreement with the exact $x$-derivative of $E(x,y)$ (Figure \ref{fig:3}d). The integrator metasurface also displays good functionality (Figures \ref{fig:3}e and \ref{fig:3}f), but it is clear that a slight overshoot is seen in the electric field away from the main lobe. This discrepancy from the exact integral is attributed to the plateau of constant reflectivity near the center of the metasurface; a conclusion that seems reasonable in light of the improved performance when decreasing $d$ (Figure \ref{fig:3}f).

\begin{figure}[H]
	\centering
		\includegraphics[width=14cm]{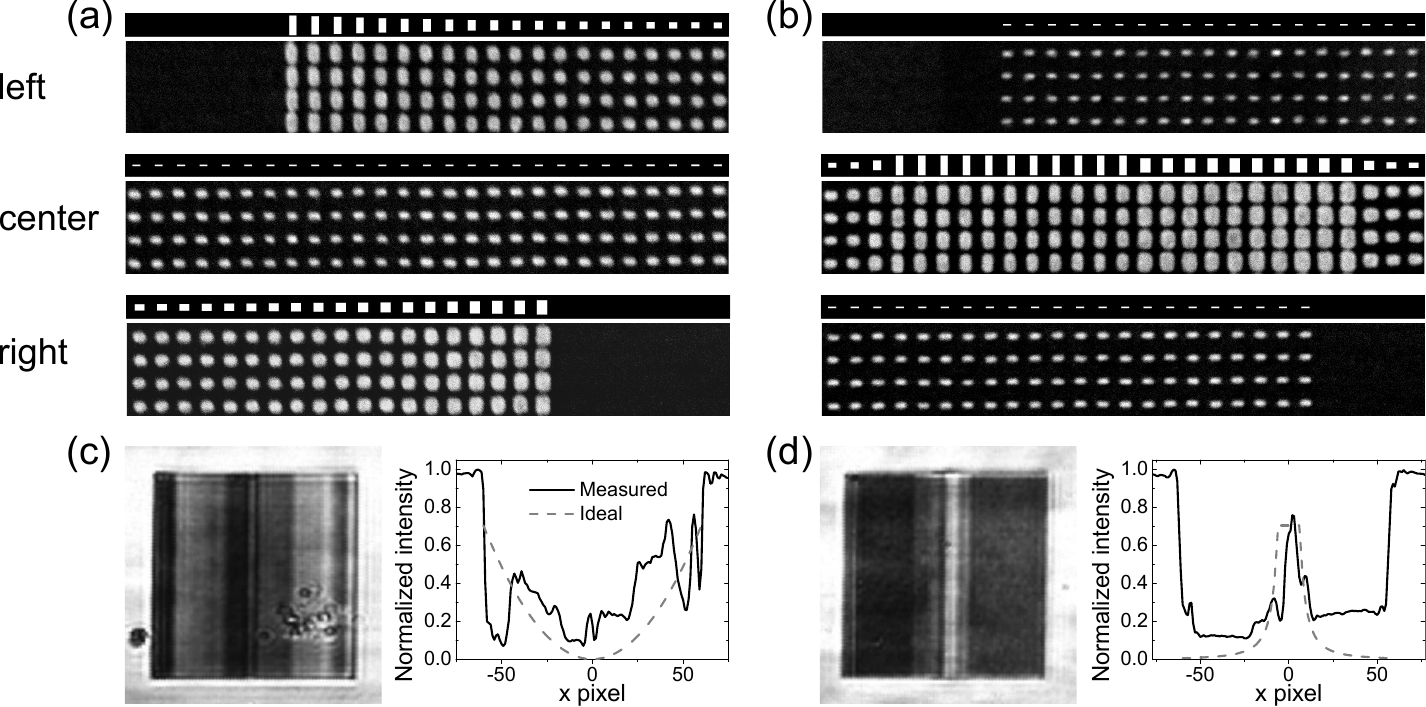}
	\caption{Fabricated metasurfaces. Scanning electron microscopy (SEM) images of the left, center, and right part of fabricated $50\times50$\,$\mu$m$^2$ (a) differentiator and (b) integrator ($d=2.5$\,$\mu$m) metasurfaces compared with similar segments of designed metasurfaces. (c,d) Bright-field images of differentiator and integrator metasurfaces, respectively, including the average normalized reflectivity along the $x$-coordinate and comparison with the expected (from ideally fabricated metasurfaces) responses.}
	\label{fig:4}
\end{figure}

\begin{figure}[H]
	\centering
		\includegraphics[width=15cm]{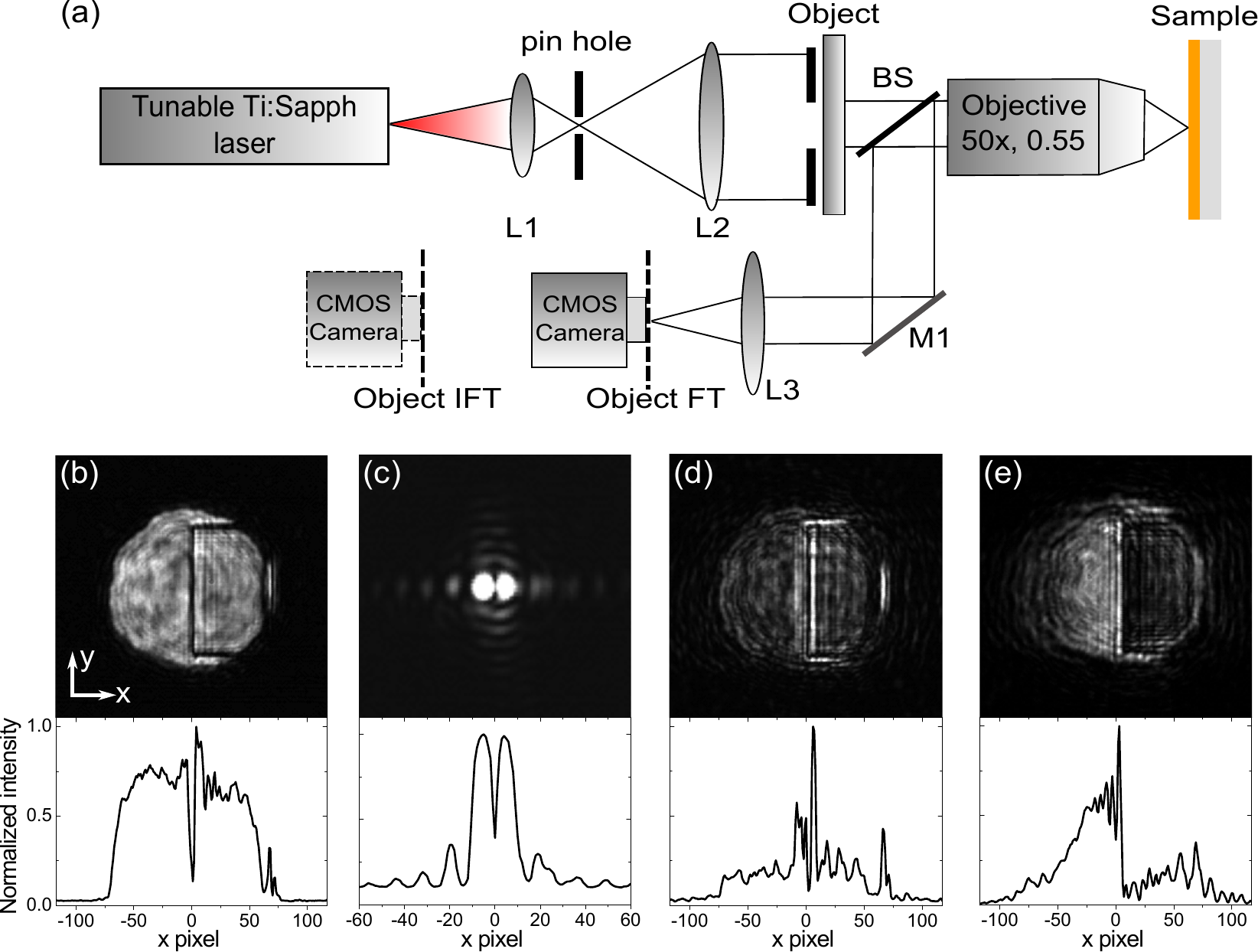}
	\caption{Experimental study of calculus metasurfaces at $\lambda=800$\,nm. (a) Sketch of experimental setup. (b) Intensity image of the phase mask under study, which consists of a rectangular patch of PMMA whose thickness is $\sim 750$\,nm so that light propagating through the patch is out-of-phase with the surrounding light. (c) Image of the intensity incident on the metasurfaces, corresponding to the Fourier transform of the object in a. Reflected far-field intensity for (d) differentiator and (e) integrator ($d=2.5$\,$\mu$m) metasurfaces. b-e show profiles of the intensity averaged along the central part in the $y$-direction as a function of the $x$-coordinate.}
	\label{fig:5}
\end{figure}

With the above numerical calculations illustrating the possibility to perform mathematical operations on incident light using GSP-based metasurfaces, we now move onto the proof-of-concept realization of differentiator and integrator metasurfaces. Figures \ref{fig:4}a and \ref{fig:4}b display representative images of the designed and lithographically fabricated metasurfaces. 
%
%
In general, we see reasonable correlation between designed and fabricated metasurfaces, though discrepancies are also clearly visible. For example, the aspect ratio of fabricated nanobricks are typically smaller than in the design due to the narrow dimension becoming too wide. As a way of probing the quality of the fabricated metasurfaces, we record the intensity of reflected light when metasurfaces are homogeneously illuminated at $\lambda=800$\,nm (Figures \ref{fig:4}c and \ref{fig:4}d). In both cases, we see noticeable discrepancies from the expected (when using ideally fabricated metasurfaces) responses. Here, it is implied that ideally fabricated metasurfaces should not only reproduce exactly designed geometrical parameters but also rely on gold exhibiting the susceptibility as tabulated\cite{johnson}. It is therefore clear that the fabrication of ideal gradient metasurfaces is a very challenging task as, for example, additional loss is often associated with grained gold nanostructures\cite{chen2}. 
Nevertheless, our previous experiments indicated certain robustness of metasurface functionalities towards imperfections\cite{pors3,pors4,pors5}, and we proceeded investigating the fabricated metasurfaces with respect to their ability to differentiate and integrate incident fields with step-like wave profiles -- a test case containing high spatial frequencies and, thus, constituting a relevant benchmark. The experimental setup is shown in Figure \ref{fig:5}a, consisting of a titanium-sapphire laser at $\lambda=800$\,nm whose beam, after proper expansion and propagation through the object, is focused onto the metasurface by a $\times50$ objective, with the reflected light being separated by the beam splitter (BS) and imaged by a lens (L3) onto the CMOS camera. It should be noted that the $\times50$ objective has the dual functionality of focusing the incident light to the size of the metasurface while working as a FT-lens. Accordingly, we can by moving the camera either image the reflected FT-field at the metasurface or the reflected twice-FT field at the back-focal plane of the objective. Note that the NA$=0.55$ of the $\times50$ objective will induce non-normal incident light, however, ensuring that only $\sim1/5$ of the entrance diameter is covered by the incident beam, we expect the influence from inclined light to be of negligible importance compared to fabrication imperfections. Finally, it is worth mentioning that the beam-shaping input object, giving the antisymmetric wave profile of the incident field, consists of a glass substrate supporting a patch of PMMA of size $0.5\times 1$\,mm$^2$ and thickness $\sim 750$\,nm, hereby ensuring that the part of light propagating through the patch will be approximately out-of-phase with the remaining light. The spatial dependence of the incident field is illustrated in Figure \ref{fig:5}b, displaying a homogeneous illumination whose spatial extent is limited by a circular aperture, with roughly half of the light propagating through the PMMA patch. The resulting FT field impinging on the metasurface (Figure \ref{fig:5}c) exhibits a clear minimum in the center, which is a signature of an antisymmetric-like wave profile with zero or small DC component ($k_x=k_y=0$), while the appearance of multiple side lobes is a consequence of the step-like wave profile. It should be noted that the intensity map in Figure \ref{fig:5}c is saturated in order to emphasize the presence of side lobes on the metasurface. Regarding the performance of the fabricated differentiator metasurface, Figure \ref{fig:5}d displays the intensity of the reflected light, demonstrating (despite a noisy background) clear peaks at the vertical edges of the PMMA patch which is coinciding with the positions of step-like variation in the incident amplitude profile. As a surprise, the metasurface, despite only intended to perform differentiation along the $x$-direction, also captures the horizontal edges of the PMMA patch; a fact related to light reflected from the surrounding gold film, as discussed below. Turning to the integrator metasurface (Figure \ref{fig:5}e), the functionality is clearly different from the differentiator metasurface, featuring a maximum in reflected intensity at the center vertical edge of the patch and a strongly asymmetric response. Ideally, the intensity should fall off in a quadratic fashion away from the vertical edge, as only observed in the left half of Figure \ref{fig:5}d, hence illustrating the non-ideal functionality of the fabricated metasurface. 

\begin{figure}[H]
	\centering
		\includegraphics[width=8.6cm]{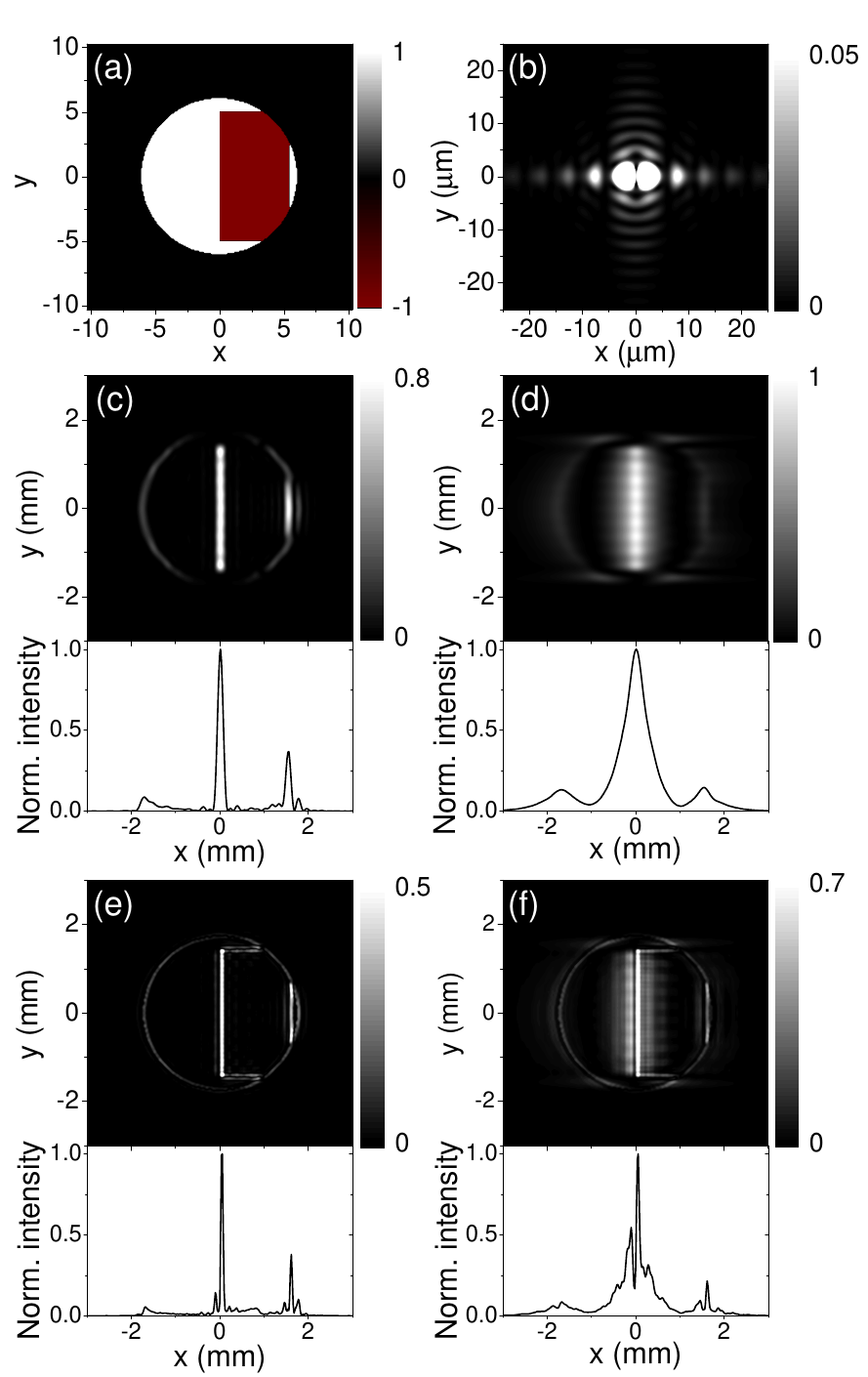}
	\caption{Numerical modeling of phase mask experiment. (a) Electric field at the output of the phase mask. (b) Intensity of the electric field impinging on the metasurfaces, corresponding to the Fourier transform of the field in a. (c,d) Intensity of reflected field from differentiator and integrator ($d=2.5$\,$\mu$m) metasurfaces, respectively, at an evaluation plane $z=10$\,mm away from metasurfaces. (e,f) are similar to c,d, but the calculations also include light reflected from the surrounding plane gold film in an area of $150\times150$\,$\mu$m$^2$. c-f show profiles of the averaged intensity along the $y$-direction as a function of the $x$-coordinate. Note that all intensity maps and profiles are normalized to one; however, the scale bars of intensity maps are chosen to better highlight weak-intensity features.}
	\label{fig:6}
\end{figure}

As a way of benchmarking the performance of fabricated metasurfaces, we resort to numerical point-dipole calculations of the intensity of the reflected light from the ideal metasurfaces designed in Figure \ref{fig:2}. The amplitude profile of the incident light used in the calculations are depicted in Figure \ref{fig:6}a and resembles the experimental situation (Figure \ref{fig:5}a). This is also evident from the FT field on the metasurface (Figure \ref{fig:6}b) which closely emulates the experimentally measured counterpart (Figure \ref{fig:5}b) with respect to intensity distribution and number of side lobes within the size of the metasurface. Figures \ref{fig:6}c and \ref{fig:6}d display the reflected far-field for differentiator and integrator metasurfaces, respectively, in which the first mentioned, as expected, produces peaks at the step-like changes in the incident amplitude along the $x$-direction, while the latter of the two, in agreement with integration along the $x$-axis, gives rise to a smooth distribution of intensity with maximum in the center and a decay towards the edges of the circular aperture. Note that the small secondary peaks in intensity at the rim of the aperture in Figure \ref{fig:6}d are a result of the approximations involved in realizing integrator metasurfaces (see, e.g., eq \ref{eq:int}). In comparison with experimental results (Figures \ref{fig:5}d and \ref{fig:5}e), it is important to note that the recorded intensity images include contributions from the weak (but many) side lobes of the incident FT field (Figure \ref{fig:5}c) that are efficiently reflected from the surrounding flat gold film and, consequently, interfere with light reflected from the metasurface. The influence of these high spatial frequencies is studied in figures \ref{fig:6}e and \ref{fig:6}f, showing the reflected far-field from an area of $150\times150$\,$\mu$m$^2$ with the $50\times50$\,$\mu$m$^2$ metasurface placed in the center. Regarding the differentiator metasurface (Figure \ref{fig:6}e), the interfering high spatial frequency components result in a narrowing of the intensity peaks and the appearance of the horizontal edges of the PMMA patch. This outcome is a consequence of the surrounding metal film functioning as a spatial high-pass filter. Since the purpose of the integrator metasurface is to suppress high-spatial frequencies, the contribution from the gold film severely degrades the performance, as evident in Figure \ref{fig:6}f. It is seen that the previously smooth and symmetric (with respect to the $x$-axis) profile (Figure \ref{fig:6}d) now shows a more erratic and asymmetric intensity distribution, though not as prominent as in the experimental case (Figure \ref{fig:5}e).

\begin{figure}[H]
	\centering
		\includegraphics[width=8.6cm]{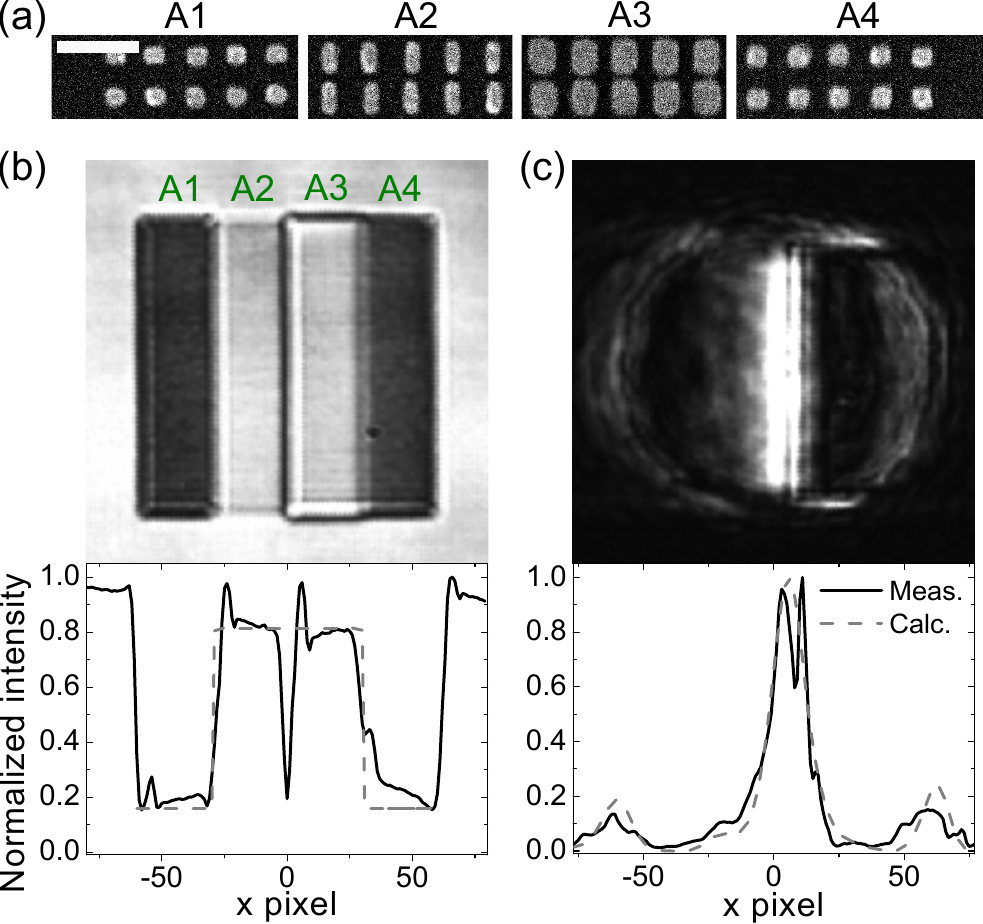}
	\caption{Poor man's integrator metasurface. (a) SEM images of the four parts constituting the $50\times50$\,$\mu$m$^2$ metasurface. Scale bar is 0.5\,$\mu$m. (b) Bright-field image of the metasurface, with the lower panel displaying the intensity averaged along the central part in the $y$-direction as a function of the $x$-coordinate. The dashed line indicates the ideal reflectivity along the $x$-direction. (c) Reflected far-field intensity for the input object of Figure \ref{fig:5}b, with the lower panel displaying the intensity averaged along the central part in the $y$-direction as a function of the $x$-coordinate. The dashed line represents the calculated intensity distribution for the ideal metasurface when neglecting the contribution from the surrounding gold film.}
	\label{fig:7}
\end{figure}

From the above results and discussion, it is clear that both types of fabricated metasurfaces show features of their designed functionality, though imperfections in the spatial variation of the reflection amplitude and phase degrades the performances, especially apparent for the integrator metasurface. The imperfections arise mainly from fabrication tolerances, but also uncertainty in the material optical properties with respect to tabulated bulk values contributes to the discrepancy between the numerical and experimental results. That said, the essence of reflective computing metasurfaces lies in realizing metasurface elements that would efficiently reflect light with a designed and \emph{equal} strength while introducing a $\pi$-phase difference. As a clear-cut example that would experimentally demonstrate this possibility with GSP-based metasurfaces, we have chosen a so-called poor man's integrator metasurface (Figure \ref{fig:7}). The ideal reflectivity from such a metasurface is indicated with a dashed line in the lower panel of Figure \ref{fig:7}b, and it should consist of four equally-sized homogeneous metasurfaces, with the two areas constituting the center part being highly reflective (A2 and A3 in Figure \ref{fig:7}b), while the outermost areas should absorb most the the incident light (A1 and A4 in Figure \ref{fig:7}b). Moreover, the two halves of the metasurface reflect light with a $\pi$-phase difference, hence making the configuration to perform as an integrator-like metasurface and not just as a low-pass filter. By careful optimization of fabricated nanobrick dimensions we succeeded in realizing a high-quality poor-man's integrator metasurface, with examples of the four nanobrick arrays constituting the metasurface imaged in Figure \ref{fig:7}a. Importantly, the two highly reflective areas of the metasurface (A2 and A3) contain nanobricks of markedly different sizes, hereby indicating (in accordance with Figure \ref{fig:2}a) a noticeable difference in the phase of the reflected light. This is indeed the case, as seen in the intensity of the reflected light from homogeneous illumination of the metasurface (Figure \ref{fig:7}b), featuring approximately the same reflectivity in the two center parts, while an approximately $\pi$-phase difference in the reflection phase is visualized by a strong dip in reflection at the border between the two center parts due to destructive interference in their responses. The outer areas of the metasurface show strongly reduced reflectivity in accordance with our design. When using the same input wave as in previous experiments, Figure \ref{fig:7}c displays the far-field intensity of the reflected light, demonstrating a fairly good agreement with the ideal (i.e., in accordance with numerical simulations) performance (dashed line). It should be elucidated that a weak dip seen at the center of the main lobe is a result of interference with high-spatial frequencies being reflected from the surrounding gold film, as discussed in relation to Figure \ref{fig:6}. Finally, it is worth noting the overall good agreement in the response from the originally designed integrator (Figure \ref{fig:6}d) and the poor man's version (Figure \ref{fig:7}c; dashed line). Realizing, however, that such approximations are typically involved in the design of holograms\cite{walter2012,ni2013}, it seems natural that similar simplifications can also be successfully applied to calculus operations.

In conclusion, we have proposed, designed, and by numerical calculations verified the first metasurface configuration that allows for mathematical operations on incident electromagnetic waves at visible wavelengths. The configuration is based on GSP-based metasurfaces\cite{nielsen,roberts2,hao,pors6,pors7,dai,li,pors5,sun2,sun,farahani,pors3,pors4,chen}, thus working in a reflection setup, which results in relatively simple designs, realizable in one step of electron beam lithography. Proof-of-concept experiments have been conducted on differentiator and integrator metasurfaces, demonstrating in both cases features of the designed functionality. We foresee that the presented approach stimulates further experiments perfecting the considered functionalities as well as new developments, including transfer to other frequency ranges and extension to other mathematical operations, such as, e.g., the second derivative for application within ultra-fast edge detection. Moreover, the possibility to control both the amplitude and phase of the reflected light with GSP-based metasurfaces may find applications within synthesis of complex wave shapes\cite{liu2014} and information storage in true (i.e., amplitude and phase modulated) holograms\cite{walter2012,ni2013}.

\section*{Notes}
The authors declare no competing financial interests.

\begin{acknowledgement}
We acknowledge financial support for this work from the Danish Council for Independent Research (the FNU project, contract no. 12-124690) and European Research Council, Grant 341054 (PLAQNAP). We would like to thank Ilya P. Radko for assistance in assembling and aligning the experimental setup. 
\end{acknowledgement}

%

\providecommand*\mcitethebibliography{\thebibliography}
\csname @ifundefined\endcsname{endmcitethebibliography}
  {\let\endmcitethebibliography\endthebibliography}{}

\newpage

\begin{tocentry}

\centering\includegraphics[width=8.7cm]{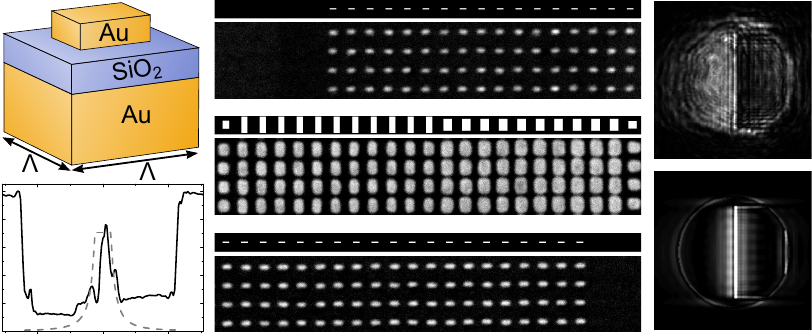}

\end{tocentry}


\pagebreak

\end{document}